\documentclass[letterpaper,12pt]{article}
\usepackage{color,amssymb,enumerate,float}
\usepackage{amsmath}
\usepackage[normalem]{ulem}

\usepackage{ifpdf}
\ifpdf
	\usepackage[pdftex]{graphicx}
	\usepackage[pdftex,unicode,implicit]{hyperref}

	\hypersetup{
  	pdftitle     = {}, 
  	pdfkeywords  = {},
  	pdfauthor    = {},
  	pdfcreator   = {pdf\LaTeXe\ with package \flqq hyperref\frqq},
  	pdfproducer  = {pdf\LaTeXe\ with package \flqq hyperref\frqq},
  	pdfpagemode  = UseNone,  
  	pdffitwindow = true,  
  	unicode      = true,
  	plainpages   = true,
  	colorlinks   = true,  
  	citecolor    = black,  
  	urlcolor     = blue, 
  	linkcolor    = black
	}

\else

  \usepackage[dvips]{graphicx}
  
\fi 

\makeatletter
\@addtoreset{equation}{section}
\makeatother

\begin{document}

\thispagestyle{empty}
	
\begin{center}
 {\bf \LARGE 
 Gauge conditions in the perturbative quantization of quadratic gravity and its quantum Lagrangian
 }
 \vspace*{15mm}
	
 {\large Jorge Bellorin}$^{1,a}$,
 {\large Claudio B\'orquez$^{2,b}$}
 {\large and Byron Droguett}$^{1,c}$
 \vspace{3ex}
 
 $^1${\it Department of Physics, Universidad de Antofagasta, 1240000 Antofagasta, Chile.}
 
 $^2${\it Facultad de Ingenier\'ia, Universidad San Sebasti\'an, Lago Panguipulli 1390, Puerto Montt, Chile.}
 
 \vspace*{2ex}
 $^a${\tt jorge.bellorin@uantof.cl,} \hspace{1em}
 $^b${\tt claudio.borquez@uss.cl,} \hspace{1em}
 $^c${\tt byron.droguett@uantof.cl}

 \vspace*{15mm}
 {\bf Abstract}
\begin{quotation}{\small}
In the perturbative formulation of the Hamiltonian of quadratic gravity, a specific gauge condition is required to hold the equivalence of the Hamiltonian equations of motion with the covariant field equations. Here we give robust evidence that this is an effect of perturbations, the exact nonperturbative equations of both formulations are equivalent, with no need of specific gauge conditions. Despite this, the required gauge condition must be imposed in the perturbative approach. We use it, together with other conditions obtained on very broad grounds, to obtain the quantum Lagrangian resulting from the integration on the canonical variables of the BFV quantization of the theory. This allow us to characterize the quantum modes from the Lagrangian perspective.
\end{quotation}
\end{center}
	
\section{Introduction}
In this letter we address a technical question about the perturbative Hamiltonian formulation of quadratic gravity, and its implications on the gauge-fixing conditions used in the quantum Lagrangian. We consider the quadratic gravity as the theory given by the most general classical Lagrangian of second order in curvature. The Lagrangian is
\begin{equation}
	\mathcal{L} = 
	\sqrt{-g^{\text{\tiny(4)}}} \left( \kappa^{-2} ( R^{\text{\tiny(4)}} - 2 \Lambda ) 
	- \alpha R^{\text{\tiny(4)}}_{\mu\nu} R^{\text{\tiny(4)}}{}^{\mu\nu}
	+ \beta {R^{\text{\tiny(4)}}}^2 \right) \,,
	\label{LagrangianR2}
\end{equation}
where, as usual, the $R^{(4)}_{\alpha\beta\mu}{}^\nu R^{(4)\,\alpha\beta\mu}{}_\nu$ term has been eliminated in favour of the two quadratic terms written in (\ref{LagrangianR2}), by using the Gauss-Bonnet topological invariant in 4D. The central quantum features of quadratic gravity are that it is a perturbatively renormalizable theory, but it has quantum states of negative norm \cite{Stelle:1976gc}.

The complete Hamiltonian formulation of quadratic gravity was developed by Buchbinder and Lyahovich \cite{Buchbinder:1987vp}, and the Hamiltonian equations of motion were shown explicitly in Ref.~\cite{Bellorin:2025kir}. A previous analysis on this can be found in Ref.~\cite{Boulware:1983yj}. There are other studies related to the Hamiltonian formulation of quadratic gravity, see for example \cite{Kaku:1982xt,Demaret:1995wp, Kluson:2013hza, Glavan:2023cuy,Salvio:2024joi,Santillan:2025qaw}. The question mentioned above consists in the equivalence between the Hamiltonian equations of motion and the covariant field equations derived from (\ref{LagrangianR2}). In the linearized version of the theory around Minkowski spacetime, a specific additional (or gauge) condition must be imposed in order to hold the equivalence (whenever general-relativity terms are active, $\kappa^{-2} \neq 0$). This was found in Ref.~\cite{Bellorin:2025kir}, and the condition is that the perturbative spatial metric must be traceless, $h_{ii} = 0$. We highlight the importance of this condition. For a theory based on general covariance, the validity of the classical Hamiltonian formulation is supported by the fact that the Hamiltonian equations of motion are equivalent to the Lagrangian ones. If some condition on the additional variables is necessary to ensure the equivalence between the equations of motion, then it is a necessary condition to hold the consistency of the whole classical Hamiltonian formulation. Moreover, the quantum theory is obtained by canonical quantization; hence the required condition is a condition for the quantum theory as well, since it requires a consistent definition of the Hamiltonian operator. Therefore, it is important to further study additional conditions of this kind in quadratic gravity.

Since the additional condition was found in the linearized theory, in this letter we investigate whether this is a true condition of the nonperturbative theory, or, instead, it is an effect of perturbations. We employ the symbolic computational tool Cadabra \cite{Peeters:2006kp}. Actually, the complete comparison between field equations for general configurations is a tremendous task, even with the computational software. Our strategy is to do the computation on manageable cases, with enough complexity to get convincing evidence about the equivalence. We perform the comparison on orthogonal configurations that, in one case are time independent, and in other case depend only on time. Our main result in this part is that the Hamiltonian equations of motion and the covariant ones are totally equivalent for these two cases, with no need of imposing specific additional conditions. These results are sufficient to allow us to conclude that the exact, nonperturbative, Hamiltonian equations of motion are equivalent to the covariant ones in the general case; the need for a specific additional condition found in Ref.~\cite{Bellorin:2025kir} is an effect of the perturbative approach.

Having clarified the perturbative origin of the additional condition, we may adopt a broad class of full gauge-fixing conditions in the framework of the Batalin-Fradkin-Vilkovisky (BFV) perturbative quantization of quadratic gravity. These gauge conditions were studied in Ref.~\cite{Bellorin:2025rtw}. Using the BFV quantization is important for the Hamiltonian formalism, since this approach extends the gauge conditions by incorporating noncanonical variables and time derivatives. Due to we use the perturbative approach, for consistency we must include the condition $h_{ii} = 0$.

From the BFV quantization we may obtain the quantum Lagrangian. To this end, we perform the integration on the canonical variables, keeping the geometrical variables of the foliation that can be identified with the components of the spacetime metric. In this way, the gauge-fixing terms in the resulting quantum Lagrangian are supported by a consistent Hamiltonian formalism. We obtain them under very broad assumptions, as we have commented. We emphasise the necessity of imposing the gauge condition $h_{ii} = 0$, in the case of perturbations around Minkowski spacetime. We also want to emphasise the importance of starting with the Hamiltonian formalism to connect with the quantum Lagrangian, which is widely the most used approach. As we commented above, the Hamiltonian operator is a necessary ingredient of the quantum theory, and the canonical quantization requires the identification of the canonically conjugate pairs. Therefore, it is important to attend this scheme of quantization. There are various related studies about gauge-fixing conditions, the degrees of freedom and the form of the propagators in the quantum Lagrangian of quadratic gravity, see for example \cite{Antoniadis:1986tu,Alves:2025zkk,Hell:2023mph,Edery:2019bsh,Chakraborty:2020ktp}.

We obtain the propagators of the quantum modes from the quantum Lagrangian. The results are essentially the same of the Hamiltonian formalism \cite{Bellorin:2025rtw}, and are in agreement with the original modes found by Stelle \cite{Stelle:1976gc}. The distribution of the physical modes is different to the analysis of Stelle, since we end with an Arnowitt-Deser-Misner (ADM) parameterization of the spacetime metric \cite{Arnowitt:1962hi}. We obtain two massive scalars, one massless and one massive tensor, and one massive vector. One of the massive scalars, the massive tensor and the massive vector have negative quantum norm. We present some discussion about the interpretation of the massive vector, which could be different.
 
\section{The matching with the covariant field equations}
The covariant field equations, derived from (\ref{LagrangianR2}), take the form $H_{\mu\nu} = 0$, where\footnote{We denote four-dimensional spacetime tensors with a label $^{(4)}$. We use the shorthands $\nabla^{(4)}_{\alpha\beta\cdots} = \nabla^{(4)}_\alpha \nabla^{(4)}_\beta\cdots$, and similarly with the spatial derivative.}
\begin{equation}
\begin{split}
	H_{\mu\nu} \equiv &
	- \kappa^{-2} R^{\text{\tiny(4)}}_{\mu \nu} 
	+ 2 \alpha R^{\text{\tiny(4)}}_{\alpha \mu} 
	R^{\text{\tiny(4)}}{}^{\alpha}{}_{\nu} 	 
	- 2 \beta R^{\text{\tiny(4)}} R^{\text{\tiny(4)}}_{\mu \nu} 
	- 2 \alpha \nabla^{\text{\tiny(4)}}_{\alpha(\mu}     
	R^{\text{\tiny(4)}}_{\nu)}{}^\alpha 
	+ \alpha \nabla^{\text{\tiny(4)}}_{\alpha}{}^{\alpha} 
	R^{\text{\tiny(4)}}_{\mu \nu} 
	\\&
	+ 2 \beta \nabla^{\text{\tiny(4)}}_{\mu \nu} R^{\text{\tiny(4)}}	
	+ \frac{1}{2} g^{\text{\tiny(4)}}_{\mu \nu} \left( 
	{\kappa}^{-2} ( R^{\text{\tiny(4)}} - 2\Lambda )    
	- \alpha R^{\text{\tiny(4)}}_{\alpha\beta} 
	R^{\text{\tiny(4)}}{}^{\alpha\beta}
	+ \beta {R^{\text{\tiny(4)}}}^{2}
	+ \upsilon_4 \nabla^{\text{\tiny(4)}}_\alpha{}^\alpha 
	R^{\text{\tiny(4)}} \right)  \,.
\end{split}
\label{covEOM}
\end{equation}

The Hamiltonian formalism of quadratic gravity is based in the ADM parametrization of the spacetime metric, which uses the fields $N, N^i, g_{ij}$. We raise and lower spatial indices with the spatial metric $g_{ij}$. We employ the standard notation of the Riemannian geometry, as the symbols $\nabla_i$ and $R_{ijk}{}^{l}$, for the geometrical objects referred to $g_{ij}$. As usual, we denote by $\pi^{ij}$ the canonical momentum conjugate to $g_{ij}$. As a geometrical variable, the extrinsic curvature tensor of the spatial hypersurfaces is defined by
\begin{equation}
	K_{ij} = \frac{1}{2N} \left( \dot{g}_{ij} - 2 \nabla_{(i} N_{j)} \right) \,.
	\label{K}
\end{equation}
In the Hamiltonian formulation of quadratic gravity, the variable $K_{ij}$ is taken as an independent canonical variable, incorporating its conjugate momentum $P^{ij}$. With the aim of abbreviating the notation, we introduce the (nongeometrical) index $A = 0,i\,$, such that $N$ and $N^i$ can be grouped in one single symbol: $N^A = N,N^i$, and similarly $P_A = P_N, P_i$. We list several combinations we use in the Hamiltonian formulation:
\begin{equation}
	\begin{split}
		&
		\upsilon_m = \alpha - m\beta \,, \quad  m=1,2,3\ldots, \quad
		\gamma = ( 8 \alpha \upsilon_3 )^{-1} \,,
		\\&
		V_i = \nabla^j K_{ij} - \nabla_i K  \,,
		\quad
		L = K_{ij} K^{ij} - K^2 \,,
		\\&
		J_{ijkl} = 2 K_{ik} K_{jl} - K_{ij} K_{kl} \,,	
		\quad J_{ij} = g^{kl} J_{ijkl} \,,
		\\&
		Q^{i j} = 
		P^{i j} + 2 \sqrt{g} \left( \alpha  R^{ij} + g^{ij} X \right)  
		\,, 
		\\& 
		\tilde{Q}_{ij} = 
		-(2\alpha)^{-1} Q_{ij} + \gamma \upsilon_4 g_{ij} Q \,,
		\quad
		\hat{Q}_{ij} = \tilde{Q}_{ij} + \sqrt{g} R_{ij} \,,
		\\&
		V =  
		\sqrt{g} \left( 
		- \kappa^{-2} ( L + R - 2\Lambda ) + \alpha ( L^2 + R_{ij} R^{ij} - 2 V_i V^i ) - \beta ( L + R )^2 \right)
		\,,
		\\&
		X = - \kappa^{-2} + \upsilon_2 L - 2 \beta R \,, 
		\\& 
		Y = ( 2 \upsilon_3 )^{-1}
		\left( 
		- \beta P 
		+ 2 \alpha \sqrt{g} \left( 
		\kappa^{-2} - \beta ( L - R ) 
		\right) \right) \,, 
		\\&
		Z = ( 2 \upsilon_3 )^{-1}
		\left(
		\upsilon_2 P
		- 2 \alpha \sqrt{g} \left(
		\kappa^{-2} 
		+ \upsilon_4 ( L - R ) 
		\right) \right) \,.
	\end{split}
\end{equation}
We denote traces by eliminating the indices: $K = g^{ij} K_{ij}$, $\pi = g_{ij} \pi^{ij}$, $P = g_{ij} P^{ij}$, $Q = g_{ij} Q^{ij}$, and $ \tilde{Q} = g^{ij} \tilde{Q}_{ij}$.

The Hamiltonian approach for this theory must be separated for the several cases of values of the coupling constants \cite{Buchbinder:1987vp}, with changes in the number of physical degrees of freedom for each case. The case that we can consider as the generic one requires the conditions $\alpha \neq 0$ and $\beta \neq \alpha / 3$, which we assume in this study. When these conditions hold, the Legendre transformation can be done on a straightforward way. All the constraints of the generic case are of first class. It has eight independent physical degrees of freedom; among them there are the negative-norm states. The bulk part of the Hamiltonian is a sum of constraints,
\begin{equation}
 H = \int d^3x N^A T_A \,.
 \label{primaryHamiltomian}
\end{equation}
In the above, we denote the four constraint as $T_A = 0$, where
\begin{eqnarray}
 &&
 T_0 = 
   \frac{1}{2\sqrt{g}} \tilde{Q}_{ij} Q^{ij}
 + \left( \nabla_{ij} + J_{ij} \right) P^{ij} 
 + 2 K_{ij} \pi^{ij} + V \,,
 \label{T0}
 \\&& 
 T_i =
 - 2 g_{ij} \nabla_k \pi^{jk} 
 - 2 \nabla_k ( K_{ij} P^{jk} ) 
 + \nabla_i K_{jk} P^{jk}  \,.
 \label{Ti}
\end{eqnarray}
The Hamiltonian equations of motion of quadratic gravity are given by
\begin{eqnarray}
 \dot{g}_{ij} &=& 
 2 N K_{i j} + 2 \nabla_{(i} N_{j)} \,,
 \label{dotg}
 \\ 
 \dot{\pi}^{ij} &=& 
 \frac{N}{\sqrt{g}} \left( 
 \frac{1}{4} g^{i j} Q^{kl} \tilde{Q}_{k l}  
 + \frac{1}{2\alpha} Q^{ik} Q^j{}_k
 - \gamma\upsilon_4  Q Q^{i j}
 \right) 
 \nonumber \\&&
 - \alpha \left( 
   2 \nabla^{k(i} ( N \hat{Q}^{j)}{}_k ) 
 - \nabla^2 ( N \hat{Q}^{ij} ) 
 - g^{i j} \nabla^{kl} ( N \hat{Q}_{kl} )  
 \right)
 \nonumber\\&&
 + \left( \nabla^{ij} - g^{i j} \nabla^2 - R^{i j} \right) ( N Y )
 - \nabla_{k} ( P^{k(i} \nabla^{j)}N )
 + \frac{1}{2} \nabla_{k} ( P^{i j} \nabla^{k} N  )
 \nonumber \\&&
 + N \Big( 
   P^{k l} J_{kl}{}^{ij} 
 + Z ( K K^{i j} - K^{ik} K^j{}_k )
 + 2 X \tilde{Q}^{ij} 
 + 4 \alpha  R^{k(i} \tilde{Q}^{j)}{}_k 
 + 2 \alpha \sqrt{g} R^{ik} R^j{}_k
 \nonumber\\&&
 - 2 \alpha \sqrt{g} V^{i} V^{j}  
 - \frac{1}{2} g^{i j} \left( 
   2 X \tilde{Q} 
 + 2 \alpha R^{kl} \tilde{Q}_{kl} 
 + V \right)
 \Big) 
 \nonumber \\&&
 - 2 \alpha \sqrt{g} \left(
   K^{i j} \nabla_{k} ( N V^{k} )   
 - N V^{k} \nabla_{k}{K^{i j}} 
 - 2 K^{k(i} \nabla^{j)} ( N V_{k} )
 + g^{i j} \nabla_{k}( N V_{l} K^{k l} ) \right)
 \nonumber\\&&
 + N^k \nabla_{k} \pi^{i j} 
 - 2 \pi^{k(i} \nabla_{k} N^{j)} 
 + \pi^{i j} \nabla_{k} N^k
 \,,
 \label{dotpi}
 \\
 \dot{K}_{i j} &=&   
 \frac{N}{\sqrt{g}} \tilde{Q}_{ij}
 + \left( \nabla_{i j} + J_{ij} \right) N
 + N^k \nabla_{k} K_{i j} + 2 K_{k(i} \nabla_{j)} N^{k} \,,
 \label{EOMdotK}
 \\
 \dot{P}^{i j} &=& 
 N \left(
 - 2 \pi^{i j} 
 + P^{i j} K - 4 P^{k (i} K^{j)}{}_{k} + Z K^{i j}
 + g^{i j} ( P^{k l} K_{k l} - Z K ) 
 \right)  
 \nonumber  \\&&
 - 4 \alpha \sqrt{g} \left( \nabla^{(i} ( N V^{j)} )
 -  g^{i j} \nabla_{k} ( N V^{k} ) \right)
 + N^k \nabla_{k} P^{i j} 
 - 2 P^{k(i} \nabla_{k} N^{j)} 
 + P^{i j} \nabla_{k} N^k  \,.
 \nonumber\\&&
 \label{EOMdotP}
\end{eqnarray}

We may perform the comparison between the set of equations defined by the constraints (\ref{T0}) and (\ref{Ti}) and the Hamiltonian equations of motion (\ref{dotg}) -- (\ref{EOMdotP}), with the set of covariant field equations (\ref{covEOM}). To do this, we substitute the ADM variables in the covariant field equations. To deal with manageable cases, we perform the comparison assuming some simplifying conditions. We start by imposing a general orthogonal ansatz, defined by $ N^i = 0 $. Under this condition, we consider two different cases, which can be considered as opposite each other: the static case and the spatially independent case. 

In the static case the time derivatives of all fields vanish. From the Eqs.~(\ref{dotg}) and (\ref{EOMdotP}) we get $K_{ij} = \pi^{ij} = 0$. The Eq.~(\ref{EOMdotK}) yields
\begin{equation}
 \tilde{Q}_{ij} = - \sqrt{g} \frac{\nabla_{ij} N}{N} \,.
\end{equation}
With the conditions we have so far, the constraint $T_i$ (\ref{Ti}) is automatically solved, and the components $H_{0i}$ of the covariant field equations (\ref{covEOM}) are also automatically zero. Hence, this sector of the field equations is in agreement. We compare the time-time component $H_{00}$ with the constraint $T_0$. We pose the equality 
\begin{equation}
H_{00} = \frac{N^2}{2\sqrt{g}} T_0 \,.
\label{h00t0}
\end{equation}
By evaluating this relation, we get
\begin{equation}
 \alpha N \left(  
   \nabla^4 N
 - \nabla_{ij}{}^{ij}{N}
 + R_{i j} \nabla^{i j}{N} 
 + \frac{1}{2} \nabla_{i}{N} \nabla^{i}{ R }
  \right)
 = 0 \,.
 \label{preremaining}
\end{equation}
We may rewrite the second term as
\begin{equation}
 \nabla_{ij}{}^{ij}{N} =
 \nabla^4 N + \nabla_i R^i{}_k \nabla^k N + R^i{}_k \nabla_i{}^k N \,.
 \label{nabla4}
\end{equation}
Hence, the left-hand side of (\ref{preremaining}) is identically zero, such that the relation (\ref{h00t0}) is true. Now we move to the space-space components. We write the Eq.~(\ref{dotpi}) in the form $\dot{\pi}^{ij} = \mathcal{E}^{ij}$, and pose the relation
\begin{equation}
	H^{ij}  = \frac{1}{\sqrt{g} N} \mathcal{E}^{ij} \,.
	\label{hijpiij}
\end{equation}
Under the conditions we are assuming, the factor $\mathcal{E}^{ij}$ becomes
\begin{eqnarray*}
\mathcal{E}^{ij} &=& 
\frac{N}{\sqrt{g}} \left( 
\frac{1}{4} g^{i j} Q^{kl} \tilde{Q}_{k l}  
+ \frac{1}{2} Q^{ik} Q^j{}_k
- \frac{1}{8} Q Q^{i j}
\right) 
+ \left( \nabla^{ij} - g^{i j} \nabla^2 - R^{i j} \right) ( N Y )
\nonumber\\&&
- 2 \nabla^{k(i} ( N \hat{Q}^{j)}{}_k ) 
+ \nabla^2 ( N \hat{Q}^{ij} ) 
+ g^{i j} \nabla^{kl} ( N \hat{Q}_{kl} )  
- \nabla_{k} ( P^{k(i} \nabla^{j)}N )
+ \frac{1}{2} \nabla_{k} ( P^{i j} \nabla^{k} N  )
\nonumber \\&&
+ N \Big( 
  2 X \tilde{Q}^{ij} 
+ 4 R^{k(i} \tilde{Q}^{j)}{}_k 
+ 2 \sqrt{g} R^{ik} R^j{}_k
- \frac{1}{2} g^{i j} \left( 
2 X \tilde{Q} 
+ 2 R^{kl} \tilde{Q}_{kl} 
+ V \right)
\Big) 
\,.
\end{eqnarray*} 
The evaluation of relation (\ref{hijpiij}) yields 
\begin{equation}
\begin{split}
 &
 \alpha \left( \nabla_{k(i}{ R_{j)}{}^{k} } 
 - \nabla_{(i|k}{ R_{|j)}{}^{k} } 
  + R_{k(ij)}{}^{l} R_{l}{}^{k} 
 - R_{ik} R_{j}{}^{k}
 + N^{-1} \nabla_{(i|k|j)}{}^{k}{ N }
 - N^{-1} \nabla_{k(ij)}{}^{k}{ N } 
 \right. 
 \\&
 \left.
 - N^{-1} R_{k(ij)}{}^{l} \nabla_{l}{}^{k}{ N }
 + N^{-1} R_{k(i} \nabla_{j)}{}^{k}{ N }
 \right)
 \\&
 + \alpha N^{-1} h_{ij} \left( 
   \nabla_{kl}{}^{kl}{ N }
 - \nabla^4 N
 - R_{kl} \nabla^{kl}{ N }
 - \frac{1}{2} \nabla_{k}{ N } \nabla^{k}{ R }
 \right)
 = 0 \,.
\end{split}
\label{endij}
\end{equation}
We make some massaging: for an arbitrary tensor $X_i{}^j$,
\begin{eqnarray}
 &&
 [ \nabla_{k} \,, \nabla_{i} ] X_{j}{}^{k} =
 - R_{kij}{}^{l} X_{l}{}^{k} + R_{ik} X_{j}{}^{k} \,.
\end{eqnarray}
Hence, all the terms in the first two lines of (\ref{endij}) cancel out between them. The terms between brackets of the third line are zero again by virtue of identity (\ref{nabla4}). Therefore, the relation (\ref{hijpiij}) is valid. We then have that the covariant field equations and the set of Hamiltonian equations of motion and constraints are equivalent under the present conditions.

Next, we consider the case when the configuration is independent of the spatial point on the hypersurface, having dependence only in time. As in the previous case, the mixed components $H_{0i}$ and the constraint $T_i$ vanish. We have checked that the time-time components $H_{00}$ are equivalent to the constraint $T_{0}$, through the relation (\ref{h00t0}). For the space-space components, we extend the relation between the two formulations to the following form:
\begin{equation} 
	H^{ij}  = \frac{1}{\sqrt{g} N} ( \mathcal{E}^{ij} - \dot{\pi}^{ij} ) \,.
\label{hijpiijtime}
\end{equation}
In this case, the factor $\mathcal{E}_{ij}$ reduces to
\begin{eqnarray}
	\mathcal{E}^{ij} &=& 
	\frac{N}{\sqrt{g}} \left( 
	\frac{1}{4} g^{i j} Q^{kl} \tilde{Q}_{k l}  
	+ \frac{1}{2\alpha} Q^{ik} Q^j{}_k
	- \gamma\upsilon_4  Q Q^{i j}
	\right) 
	\nonumber \\&&
	+ N \Big( 
	P^{k l} J_{kl}{}^{ij} 
	+ Z ( K K^{i j} - K^{ik} K^j{}_k )
	+ 2 X \tilde{Q}^{ij} 
	- \frac{1}{2} g^{i j} \left( 
	2 X \tilde{Q} 
	+ V \right)
	\Big) 
	\,.
\end{eqnarray}
We have checked that the equality (\ref{hijpiijtime}) holds in this case. Therefore, both kinds of field equations are equivalent in this case as well. 

The two cases we have studied give sufficient evidence to conclude that the Hamiltonian equations of motion are equivalent to the covariant ones. Note that both cases have the trace (or the determinant) of the spatial metric $g_{ij}$ active; hence, the vanishing of this trace, would be a contradiction to the fact that we have not seen any special condition in the nonperturbative approach. Thus, there is no need for specific additional conditions in the complete, nonperturbative theory. The necessity for the additional condition found in Ref.~\cite{Bellorin:2025kir} is an effect of perturbations.

\section{BFV Quantization: integration on canonical fields}
The BFV quantization of quadratic gravity was done in Ref.~\cite{Bellorin:2025rtw}. We summarise here the main steps. The Lagrange multipliers $N^A$ associated with the first-class constraints are promoted to canonical variables. Thus, we introduce the new canonical pairs $(N,P_{N})$ and $(N^{i},\tau_i)$. We can group the additional momenta in one single object: $\pi_A = P_{N},\tau_i$. Two canonically conjugate pairs of fermionic ghosts, which we first denote by $(\eta_1^{A},\mathcal{P}^{1}_A)$ and $(\eta_2^{A},\mathcal{P}^{2}_A)$, are required.

The gauge-fixed quantum Hamiltonian is given by 
\begin{equation}
	\mathcal{H}_{\Psi}= \left\lbrace\Psi,\Omega\right\rbrace \,.
	\label{generalgfhamiltonian}
\end{equation}
$\Psi$ is the gauge fermion that must be chosen, and $\Omega$ is the generator of the Becchi-Rouet-Stora-Tyutin (BRST) symmetry. For quadratic gravity, it takes the form
\begin{equation}
	\Omega =
	\int d^3x \left(
	T_A\eta_1^{A} + \pi_A\eta_2^{A}
	- \eta_1^i\partial_i \eta_1 \mathcal{P}^1
	- \eta_1^i\partial_i\eta_1^j\mathcal{P}_j^1
	\right) \,.
\end{equation}
The path integral becomes
\begin{equation}
	\begin{split} 
		&
		Z = 
		\int\mathcal{D} \mathcal{V} 
		\exp \left[
		i \int dt d^{3}x
		\Big(
		\pi^{ij} \dot{g}_{ij} 
		+ P^{ij} \dot{K}_{ij}
		+ \pi_A \dot{N}^{A}
		+ \mathcal{P}_{A}^{1}\dot{\eta}^{A}_{1}
		+ \mathcal{P}_{A}^{2}\dot{\eta}^{A}_{2}
		- \mathcal{H}_\Psi\Big) 
		\right] \,,
		\\&
		\mathcal{D}\mathcal{V} =
		\mathcal{D} g_{ij} \mathcal{D} \pi^{ij} 
		\mathcal{D}K_{ij}\mathcal{D}P^{ij}\mathcal{D}N^A \mathcal{D}\pi_A 
		\mathcal{D}\eta^A_{1} \mathcal{D}\mathcal{P}^{1}_A \mathcal{D}\eta^A_{2} \mathcal{D}\mathcal{P}^{2}_A \,.
	\end{split}
	\label{Z}
\end{equation}

The BFV formalism includes an appropriate form of the gauge fermion, which is
\begin{equation}
	\Psi = \mathcal{P}_A^{1}N^{A} + \mathcal{P}_A^{2}\chi^{A} \,,
\end{equation}
where the equations:
\begin{equation}
	\dot{N}^A - \chi^A = 0
	\label{bfvgaugegeneral}
\end{equation}
are the conditions on the arbitrary functions to be effectively imposed. To fix $\chi^A = \chi,\chi^i$ we employ perturbative variables, which are defined by
\begin{eqnarray}
	& g_{ij} = \delta_{ij} + h_{ij} \,,
	\quad
	N = 1 + n \,,
	\quad
	N^i = n^i \,, \quad
	P^{ij} = 2\kappa^{-2} \delta^{ij} + p^{ij}  \,,
\end{eqnarray}
and for the rest of fields, we keep the original notation\footnote{We use the three-dimensional transverse-longitudinal decomposition for tensors: if $\sigma_{ij}$ is a symmetric spatial tensor, we decompose it in the way
\begin{equation*}
		\sigma_{ij} = 
		\partial_{ij} \Delta^{-1} \sigma^L
		+ \partial_{(i} \sigma_{j)}^L
		+ \frac{1}{2} ( \delta_{ij} - \partial_{ij} \Delta^{-1} ) \sigma^T  
		+ \sigma_{ij}^{T} \,,
		\label{decomposition}
\end{equation*}
where $\Delta = \partial_{kk}$, with the conditions: $\sigma_{kk}^T = \partial_k \sigma_{kj}^T = \partial_k \sigma_k^L = 0$. For a spatial vector $\xi_i$, we use $	\xi_i = \partial_i \xi^L + \xi_i^T $, with $\partial_k \xi_k^T = 0$.}.

To obtain the form of the factors $\chi^A$ we assume several conditions that can be considered of broad scope. The $\chi^A$ are linear on the fields and are based on the functional condition (\ref{bfvgaugegeneral}). We assume that they do not include dimensional coupling constants. As a consequence, $\chi^A$ do not depend on time derivatives, and the possible combinations of fields arising in them are restricted by dimensionality. We also assume that $\chi^A$ are local expressions; hence only the identity and spatial derivatives can operate on the fields. We want to preserve the symmetry of spatial rotations. We do not consider any dependence of $\chi^A$ on ghost fields. Finally, for consistency, the vanishing of the trace of the spatial metric, 
\begin{equation}
 h^L = - h^T 
 \label{traceless}
\end{equation} 
must be imposed as part of the additional conditions. We find the set of additional conditions that satisfy these requirements:
\begin{eqnarray}
	&&
	\chi =
	\sigma_1\Delta n^L+\sigma_2(K^T+K^L) \,,
	\label{gauechi}
	\\&&
	\chi^L = 
	\frac{1}{\epsilon} ( h^L + h^T ) \,,
	\label{chiL}
	\\&&
	\chi_i^T = 
	\sigma_3 \Delta h_i^L \,,
	\label{gaugechiiT}
\end{eqnarray}
where $\sigma_1,\sigma_2,\sigma_3$ and $\epsilon$ are dimensionless constants. In the limit $\epsilon \rightarrow 0$, the Eqs.~(\ref{chiL}) and (\ref{bfvgaugegeneral}) become equivalent to the desired condition (\ref{traceless}). We comment that the gauge conditions (\ref{gauechi}) -- (\ref{gaugechiiT}) preserve the fourth order in derivatives of the theory, and, equivalently, the UV $k^{-4}$ scaling of the propagators, as we shall see. In other studies, gauge conditions with orders in derivatives higher than the one of  (\ref{gauechi}) -- (\ref{gaugechiiT}) have been considered, see for example Ref.~\cite{Teyssandier:1989dw}. From the analysis of Ref.~\cite{Bellorin:2025rtw}, we know that a convenient choice for the constants is $\sigma_1 = 0$, $\sigma_2 = 1$, and $\sigma_3 = \frac{1}{2}$, which we adopt from now on. This choice leads to propagators with the structure of the Klein-Gordon propagator, valid in the ADM frame, for the fields of interest. 

After the additional conditions (\ref{gauechi}) -- (\ref{gaugechiiT}), the explicit Hamiltonian (\ref{generalgfhamiltonian}) and the path integral (\ref{Z}) can be computed at second order in perturbations.
The resulting expressions are very large, they can be found in Ref.~\cite{Bellorin:2025rtw}. 

We proceed with the integration on canonical variables in the BFV path integral (\ref{Z}). In all steps, we discard factors that are independent of the quantum fields. We make a convenient change of notation on the BFV ghosts, which was used in the original BFV paper. The change is
\begin{equation}
\begin{split}
	&	
	\eta_{1}^{A} = 
	(\eta_{1},\eta_{1}^{i}) = (C,C^{i}) \,,
	\quad
    \eta_{2}^{A} = ( \eta_{2},\eta_{2}^{i} ) = (\mathcal{P},\mathcal{P}^{i}) \,,
	\\&
	\mathcal{P}_{A}^{1} = ( \mathcal{P}^{1},\mathcal{P}^{1}_i ) = (\bar{\mathcal{P}},\bar{\mathcal{P}}_i)\,,
	\quad
	\mathcal{P}_{A}^{2} = ( \mathcal{P}^{2},\mathcal{P}^{2}_i ) = 
	( \bar{C},\bar{C}_i )\,.
\end{split}
\end{equation}
The (Grassmannian) Gaussian integration on the ghost fields $\mathcal{P}$, $\bar{\mathcal{P}}$, $\mathcal{P}^i$, $\bar{\mathcal{P}}_i$ essentially leads to kinetic terms for the $C$, $\bar{C}$, $C^i$, $\bar{C}_i$ ghosts with quadratic dependence on time derivatives. The BFV canonical Lagrangian depends linearly on the components of the momentum $\pi^{ij}$. The integration on these components leads to Dirac's deltas, like $	\delta\left( \dot{h}_{ij}^{T} - 2 K_{ij}^{T} \right) $, each one involving one component of $K_{ij}$. Hence, the four Dirac's deltas coming from the integration on the components of $\pi^{ij}$ can be used to fix the integration on the components of $K_{ij}$, recovering in this way its geometrical definition (\ref{K}). The canonical Lagrangian also depends linearly on $P_N$ and $\tau_i$. The integration on these fields leads to the following Dirac's deltas: 
\begin{equation}
	\delta
	\left(
	\dot{n}-(K^T+K^L)
	\right)
	\times
    \delta
	\left(
	\dot{n}^{iT}
	-\frac{1}{2} \Delta h^{iL}
	\right)
	\times
	\delta
	\left( -\dot{n}^L+\epsilon^{-1}(h^T+h^L) \right)
	\,.
	\label{deltas}
\end{equation}
These deltas impose the gauge-fixing conditions (\ref{gauechi}) -- (\ref{gaugechiiT}), and become part of the quantum measure.

The last steps of integration are Gaussian integrations on the components of $P^{ij}$. After this, the quantum Lagrangian, separated for the several modes in the way $\mathcal{L} =  \mathcal{L}_{\text{scalar}} + \mathcal{L}_{\text{vector}} + \mathcal{L}_{\text{tensor}} + \mathcal{L}_{\text{ghost}}$, takes the form:
\begin{eqnarray}
	\mathcal{L}_{\text{scalar}}
	&=&
	- \frac{1}{8} h^T \left( ( 3\alpha - 8\beta) \Box + \kappa^{-2} \right) \Box h^T
	+ \frac{1}{2} h^T \left( \upsilon_4 \Box + \kappa^{-2} \right)
	\left( \ddot{h}^L - 2 \Delta\dot{n}^L - 2 \Delta n \right)
	\nonumber \\ && 
	- \upsilon_2 \left( 
	  \frac{1}{2} h^L\partial_t^4 h^L
	- 2 h^L\Delta \ddot{n}
	- 2 h^L\Delta\partial_t^3n^L 		  
	- n^L\Delta^2 \ddot{n}^L
	+ 2 n\Delta^2 n
	- 4 n^L\Delta^2 \dot{n}
	\right) \,,
	\nonumber \\
	\label{Lscalar}
\end{eqnarray}
\begin{eqnarray}
	\mathcal{L}_{\text{vector}}
	&=&
	- \frac{1}{8} h_i^L \left( \alpha \Box - \kappa^{-2} \right) \Delta \ddot{h}_i^L
	+ \frac{1}{2} h_i^L \left( \alpha \Box - \kappa^{-2} \right)  \Delta \dot{n}_i^T
	+ \frac{1}{2}  n_i^T \left( \alpha \Box - \kappa^{-2} \right) \Delta n_i^T \,,
	\nonumber \\
	\label{Lvector}
\end{eqnarray}
\begin{eqnarray}
	\mathcal{L}_{\text{tensor}}
	&=&
	- \frac{1}{4} h_{ij}^{T} \left( 
	\alpha \Box - \kappa^{-2} \right) \Box h_{ij}^{T} \,,
	\label{Ltensor}
\end{eqnarray}
\begin{eqnarray}
	\mathcal{L}_{\text{ghost}}
	&=&
	- \bar{C} \Box C - \bar{C}^L\Delta \ddot{C}^L
    +\bar{C}^T_k \Box C_k^T
    \,,
\end{eqnarray}
where $ \Box = - \partial^2_t + \Delta $. 

To complete the quantum Lagrangian, we consider the gauge-fixing Dirac's deltas (\ref{deltas}). They can be managed within the standard Faddeev–Popov procedure. We replace them by averaged conditions by means of the following gauge-fixing terms in the Lagrangian:
\begin{eqnarray}
	&&
	\mathcal{L}_{\text{gf}} =
	\frac{1}{2\xi_1} \left(\dot{n}^{i^T}-\frac{\Delta h^{iL}}{2}
	\right)^2
	+
	\frac{1}{2\xi_2} \left(
	\dot{n}-\frac{\dot{h}^T+\dot{h}^L}{2}+\Delta n^L \right)^2
	+ \frac{1}{2\xi_3} \left(-\dot{n}^L+\epsilon^{-1}(h^T+h^L)\right)^2
    \,.
    \nonumber \\
    \label{Lgf}
\end{eqnarray}
We note that in the limit $\epsilon \rightarrow 0$, the constant $\xi_3$ is spurious.

Having obtained the quantum Lagrangian is important in many aspects. For example, we may discuss the limit of general relativity, which is the central limiting case in any theory of quantum gravity.
As we have commented, the classical Hamiltonian formulation of quadratic gravity exhibits changes in the dimensionality of the physical phase space, as we vary the values of the coupling constants. Here we have taken the generic case; but in this case the Hamiltonian is not smoothly connected to general relativity, since we can not turn off the constant $\alpha$. Hence, it is not obvious how to make a consistent smooth connection with general relativity in the Hamiltonian language. After the integration, this obstruction disappears: the $\alpha,\beta \rightarrow 0$ limit on the factors in (\ref{Lscalar}), (\ref{Lvector}), and (\ref{Ltensor}) leads to the form
\begin{equation}
	\begin{split}
		\mathcal{L}
		=&
		\frac{\kappa^{-2}}{2} \left( - \frac{1}{4} h^T \Box h^T
		+ h^T \left( \ddot{h}^L
		- 2 \Delta \dot{n}^L
		- 2 \Delta n \right)
		+ \frac{1}{4} h_i^L \Delta \ddot{h}_i^L
		\right.
		\\
		& \left.
		- h_{i}^{L}\Delta \dot{n}_{i}^{T}
		- n_i^T \Delta n_i^T
		+ \frac{1}{2} h_{ij}^{T} \Box h_{ij}^{T}
		\right)
		\,.
	\end{split}
\end{equation}
This is the Lagrangian of general relativity in ADM variables, expanded up to second order in perturbations. Therefore, the quantum Lagrangian defined by (\ref{Lscalar}) -- (\ref{Lgf}) admits the soft limit of general relativity, under the gauge conditions we are considering.

\section{The quantum modes}
To present the independent degrees of freedom, it is sufficient to show the propagators of the components of the spatial metric. In Fourier space, defining $k^2 \equiv \omega^2 - \vec{k}^2$, we introduce the following factors that have the form of massless and massive Klein-Gordon propagators:
\begin{equation}
	\Pi_0 = \frac{i}{k^2} \,,
	\quad
	\Pi_1 = \frac{i}{k^2 - \kappa^{-2} / \alpha } \,, 
	\quad
	\Pi_{2} = \frac{i}{ k^2 + \kappa^{-2} / ( 2 \upsilon_3 ) } \,.
\end{equation}
They are valid in the frame defined by the ADM parametrization of the metric. By imposing the constraints: $\kappa^{-2} \geq 0$, $\alpha \geq 0$ and $\beta \geq \alpha / 3$, we ensure that $\Pi_1$ and $\Pi_2$ have real pole solutions for the frequency $\omega$. To fulfil the conditions of the transverse-longitudinal decomposition, we introduce the orthogonal basis $\{ e_i^{x} \}$, $\vec{k}_i e^x_i = 0$, $x = 1,2$ in the subspace of transverse vectors. In the subspace of transverse-traceless tensors, we introduce the orthogonal basis $\{ e^X_{ij} \}$, $e^X_{ij} = e^X_{ji}$, $\vec{k}_i e^X_{ij} = 0$, $e^X_{ii} = 0$, $ X=1,2$. With these bases we can write  $V_i^T = V_x^T e^x_i$ and $T^T_{ij} = T^T_X e^X_{ij}$. Since the components in these bases are unconstrained objects, we can obtain their propagators by direct inversion.

Considering the complete quantum Lagrangian, the (diagonal) propagators, after taking the limit $\epsilon \rightarrow 0$, are
\begin{eqnarray}
	&&
	\left\langle h^T h^T \right\rangle =
	16 i \gamma \upsilon_2 \Pi_1 \Pi_2 \,,
	\label{propagatorhT}
	\\&&
	- \vec{k}^2 \left\langle h_{x}^{L} h_{y}^{L} \right\rangle
	=
	- 4 \left( \alpha^{-1} \omega^2 
	- i \xi_1 \vec{k}^2 \Pi_1^{-1} \right) \Pi_0^2\Pi_1 \delta_{xy} \,,
	\label{propagatorhil}
	\\&&
	\left\langle h_{X}^{T} h_{Y}^{T} \right\rangle
	=
	2i \alpha^{-1} \Pi_0\Pi_1 \delta_{XY} \,.
	\label{propagatorn}
\end{eqnarray}
In addition, we show the ghost propagators:
\begin{equation}
	\left\langle \bar{C} C \right\rangle
	=
	\Pi_0\,,
	\quad
	-\vec{k}^{2} \left\langle \bar{C}^L C^L \right\rangle = 0\,,
	\quad
	\left\langle \bar{C}_x^T C_y^T \right\rangle
	= \Pi_0 \delta_{xy} \,.
\end{equation}
The propagators of $h^T$ and $h^{T}_{X}$, assuming $\kappa^{-2} > 0$, can be separated as follows:
\begin{eqnarray}
	&&
	\left\langle h^T h^T \right\rangle =
	\frac{ 4 }{ 3 \kappa^{-2} } \left( \Pi_2  - \Pi_1 \right) \,,
	\label{propsepartescalar}
	\\&&
	\left\langle h_{X}^{T} h_{Y}^{T} \right\rangle
	=
	\kappa^{-2} \left( \Pi_0 - \Pi_1 \right) \delta_{XY} \,.
\end{eqnarray}

In the scalar sector there are two massive modes, with the poles of $\Pi_1$ and $\Pi_2$. The negative sign in front of $\Pi_1$ in (\ref{propsepartescalar}) indicates that this corresponds to a one-particle state of negative norm. In the transverse-traceless tensorial sector there are two modes. The massless tensor $\Pi_0$ has the right sign, whereas the massive one $\Pi_1$ has a negative sign. The physical mode in $h^L_x$ is represented by the propagator $\Pi_1$ (for simplicity, we discard the terms proportional to the Faddeev-Popov constants $\xi_1$ in this discussion). The factor $\omega^2 \Pi_0^2 $ in front of the propagator (\ref{propagatorhil}) arises as a consequence of the ADM field operator we use. In Ref.~\cite{Bellorin:2025rtw}, it is indicated that this physical mode arises on a clean way in the propagator of $K_{i}^L$. That paper used the Hamiltonian formalism, where $K_{ij}$ and $h_{ij}$ are independent field variables. Here, the interpretation is that the time derivative of $h_{i}^L$, combined with other fields that compose the geometrical definition of $K_{i}^L$ (\ref{K}), gives the propagator on a simpler form, such that the factor $\omega^2 \Pi_0^2$ is not associated with independent physical modes. Moreover, the sign depends on what one takes as the fundamental field operator, since the propagator $- \vec{k}^2 \left\langle K_{x}^{L} K_{y}^{L} \right\rangle = \alpha^{-1} \Pi_{1} \delta_{xy}$ is positive \cite{Bellorin:2025rtw}. 

In the limit $\alpha,\beta = 0$, the masses of the modes of $\Pi_1$ and $\Pi_2$ become infinite, such that these two modes decouple, and we get the only two physical massless modes of general relativity contained in $h^T_X$.

In the case $\kappa^{-2} = 0$, the three propagators $\Pi_0$, $\Pi_1$ and $\Pi_2$ become equal, and all the eight physical modes of the theory become massless.

As a final comment, we note that we can introduce some changes in the dependence on canonical momenta by using more general gauge-fixing conditions. This is an issue raised by the original question about the Ostrogradsky's instability that is behind this theory. The Ostrogradsky's instability is manifested in the lacking of the quadratic dependence on part of the canonical momenta, having linear dependence instead. But, at least for some variables, one can introduce the quadratic dependence by means of the gauge-fixing conditions. It is interesting to explore this possibility. Moreover, in order to introduce canonical momenta in the gauge-fixing conditions without introducing dimensional constants in them, it is necessary to use nonlocal operators. For this reason, this case is out of our previous considerations that led to the conditions (\ref{gauechi}) -- (\ref{gaugechiiT}). Explicitly, the gauge-fixing conditions can be extended to the nonlocal conditions:
\begin{eqnarray}
	&&
    \chi =
	\sigma_1\Delta n^L + \sigma_2(K^T+K^L)
	+ \sigma_4 \Delta^{-1}(\pi^T+\pi^L)
	+ \sigma_5 \left(-\Delta\right)^{-1/2}\tau^L
	+ \sigma_6 \Delta^{-1}P_N
	\hspace*{1em}
	\label{nolocalgauechi}
	\\ &&
	\chi^L = 
	\frac{1}{\epsilon} ( h^L + h^T )
	\,,
	\label{nolocalchiL}
	\\ &&
	\chi_i^T = 
	\sigma_3 \Delta h_i^L
	+\sigma_7 \left(-\Delta\right)^{-1/2}\pi_i^L
	+\sigma_8 \Delta^{-1}\tau_k^T
	\,,
	\label{nolocalgaugechiiT}
\end{eqnarray}
where $\sigma_1,\ldots,\sigma_8$ are arbitrary dimensionless constants. In the above, we have considered the most general linear dependence on canonical momenta, allowing nonlocal operators while respecting the rest of conditions. Other fields like $\pi^{T}_{ij}$ cannot be incorporated to these conditions; hence, these components are unsensitive to the nonlocal gauge. With this nonlocal gauge-fixing conditions, we get quadratic terms like $\sim P_N^2$ and $\sim P_N \pi^T$ in the Hamiltonian $\mathcal{H}_\Psi$. However, after the integration on the canonical variables and getting the geometrical Lagrangian, this modification does not change in an essential way the results for the propagators we have found using the local gauge-fixing conditions.


\end{document}